\documentclass[fleqn]{2017SCGE}
\usepackage{xcolor}

\usepackage{ulem}

\newcommand\aap{A\&A}                % Astronomy and Astrophysics
\newcommand\apj{ApJ}                 % Astrophysical Journal
\newcommand\apjl{ApJ}                % Astrophysical Journal, Letters
\newcommand\apjs{ApJS}               % Astrophysical Journal, Supplement
\newcommand\apss{Ap\&SS}             % Astrophysics and Space Science
\newcommand\araa{ARA\&A}             % Annual Review of Astronomy and Astrophysics
\newcommand\mnras{MNRAS}             % Monthly Notices of the Royal Astronomical Society
\newcommand\nat{Nature}              % Nature
\newcommand\pasp{PASP}               % Publications of the Astronomical Society of the Pacific
\begin{document}

\ensubject{subject}

%%%%%%%%%%%%%%%%%%%%%%%%%%%%%%%%%%%%%%%%%%%%%%%%%%%%%%%
%%% Authors do not modify the information below

\ArticleType{Article}%
\SpecialTopic{SPECIAL TOPIC: }%
\Year{2017}
\Month{January}
\Vol{60}
\No{1}
\DOI{****}
\ArtNo{000000}
\ReceiveDate{}
\AcceptDate{}
%\OnlineDate{January 1, 2016}
%%%%%%%%%%%%%%%%%%%%%%%%%%%%%%%%%%%%%%%%%%%%%%%%%%%%%%%

%%% title: 
%%%   \title{title}{title for citation}
%\title[Periodic variations]{Periodic variations in the interstellar scintillation of PSRs~J0613$-$0200 and J0636+5128}
\title[Periodic scintillation variations]{Periodic interstellar scintillation variations of PSRs~J0613$-$0200 and J0636+5128 associated with the Local Bubble shell}

%%% Corresponding author
%%%   \author[number]{Full name}{{email@xxx.com}}
%%% General author: ???????
%%%   \author[number]{Full name}{}
\author[1,2]{Yulan Liu}{{liuyl@bao.ac.cn}}%
\author[3]{Robert A. Main}{}
\author[2,3,4]{Joris P. W. Verbiest}{}
\author[1,2]{Ziwei Wu}{}
\author[2,3]{Krishnakumar M. Ambalappat}{}
\author[1]{\\Jiguang Lu}{}
\author[3]{David J. Champion}{}
\author[5,6]{Ismaël Cognard}{}
\author[5,6]{Lucas Guillemot}{}
\author[3]{Kuo Liu}{}
\author[7,8]{James W. McKee}{}
\author[3,9]{\\Nataliya Porayko}{}
\author[9,10,11]{Golam. M. Shaifullah}{}
\author[5,6]{Gilles Theureau}{}

%%% Author information for page head. 

\AuthorMark{Yulan Liu}%\authorcr

%%% Authors for citation.

\AuthorCitation{Yulan Liu, Robert A. Main, Joris P. W. Verbiest, et al}

%%% Address.
%%%   \address[number]{Address, City {\rm Postcode}, Country}
\address[1]{National Astronomical Observatories, Chinese Academy of Sciences, Beijing 100101, China}
\address[2]{Fakult\"at f\"ur Physik, Universit\"at Bielefeld, Bielefeld 33501, Germany}
\address[3]{Max-Planck-Institut f\"ur Radioastronomie, Bonn 53121, Germany}
\address[4]{Florida Space Institute, University of Central Florida, Orlando FL 32826, USA}
\address[5]{Station de radioastronomie de Nan{\c c}ay, Observatoire de Paris, Nan{\c c}ay 18330, France}
\address[6]{Laboratoire de Physique et Chimie de l'Environnement et de l'Espace, Universit\'e d’Orl\'eans, Orl\'eans 45071, France}
\address[7]{E.A. Milne Centre for Astrophysics, University of Hull, Hull HU6 7RX, UK}
\address[8]{Centre of Excellence for Data Science, Artificial Intelligence and Modelling, University of Hull, Hull HU6 7RX, UK}
\address[9]{Dipartimento di Fisica ``G. Occhialini'', Universit\`a di Milano-Bicocca, Milano 20126, Italy}
\address[10]{INAF - Osservatorio Astronomico di Cagliari, via della Scienza 5, Selargius (CA) 09047, Italy}
\address[11]{INFN, Sezione di Milano-Bicocca, Piazza della Scienza 3, Milano 20126, Italy}

%\contributions{}%

%%% Abstract. 
\abstract{Annual variations of interstellar scintillation can be modelled to constrain parameters of the ionized interstellar medium. 
If a pulsar is in a binary system, then investigating the orbital parameters is possible through analysis of the orbital variation of scintillation.
In observations carried out from 2011 January to 2020 August by the European Pulsar Timing Array radio telescopes, 
PSRs~J0613$-$0200 and J0636+5128 show strong annual variations in their scintillation velocity,
while the former additionally exhibits an orbital fluctuation. Bayesian theory and Markov-chain-Monte-Carlo methods are used to interpret these periodic variations. 
We assume a thin and anisotropic scattering screen model, and discuss the mildly and extremely anisotropic scattering cases. PSR~J0613$-$0200 is best described by mildly anisotropic scattering, while PSR~J0636+5128 exhibits extremely anisotropic scattering.
We measure the distance, velocity and degree of anisotropy of the scattering screen for our two pulsars, finding that scattering screen distances from Earth for PSRs~J0613$-$0200 and J0636+5128 are 316$^{+28}_{-20}$\,pc and 262$^{+96}_{-38}$\,pc, respectively.
The positions of these scattering screens are coincident with the shell of the Local Bubble towards both pulsars.
These associations add to the growing evidence of the Local Bubble shell as a dominant region of scattering along many sightlines.}

%%% Keywords.

%\Authorfootnote
%\noindent

\maketitle

%\tableofcontents%

%%%%%%%%%%%%%%%%%%%%%%%%%%%%%%%%%%%%%%%%%%%%%%%%%%%%%%%
%%% The main text. 
%\twocolumn\onecolumn
%%%%%%%%%%%%%%%%%%%%%%%%%%%%%%%%%%%%%%%%%%%%%%%%%%%%%%%
\begin{multicols}{2}
\section{Introduction}
\label{sec:intro}
Pulsars are highly magnetized, rotating neutron stars that emit beams of electromagnetic radiation from their magnetic poles \cite{gol68} and are embedded in an extremely tenuous interstellar medium (ISM), which contains ordinary matter, relativistic charged particles known as cosmic rays, and magnetic fields \cite{fer+01}. 
After passing through the warm ionised interstellar medium (IISM), the spatially coherent electromagnetic radiation from the pulsar is distorted, which forms an interference pattern at the observer's plane. The interference pattern drifting across the line of sight causes a fluctuation in the source's observed flux density  
with observing frequency and time, which is defined as interstellar scintillation (ISS) \cite{sch68}.
ISS analysis of pulsars, thus, allows us to probe the smallest-scale distribution and inhomogeneities of the IISM \cite{bgr98}.

Scintillation analysis typically relies on measuring the two-dimensional image of the source's observed flux density 
as a function of observing time and observing frequency, called the dynamic spectrum, where the interference maxima in this dynamic spectrum are called scintles. 
To quantify the average characteristics of scintles,  Cordes \& Wolszczan (1986) \cite{cw86} defined two scintillation parameters: the scintillation bandwidth $\nu_{\rm d}$ and scintillation timescale $\tau_{\rm d}$, where $\nu_{\rm d}$ is the half-width at a half maximum along the frequency axis and $\tau_{\rm d}$ is the half-width at 1/e along the time axis in the two-dimensional auto-correlation function of the dynamic spectrum. At centimetre wavelengths, diffractive interstellar scintillation (DISS) appears as pulsar flux density variations in both time and frequency with characteristic scales $\sim$ minutes and $\sim$ MHz for pulsar timing array (PTA) pulsars, respectively \cite{lmj+16,lvm+22}. At metre wavelengths, the scintillation bandwidth is typically as small as $\sim$ kHz for nearby pulsars \cite{wvm+22}.

The effective velocity, $V_{\rm{eff}}$, of the line of sight relative to the medium (which is a linear combination of the transverse velocities of Earth, the pulsar, and the IISM) has a strong impact on scintillation parameters, particularly the scintillation timescale $\tau_{\rm d}$. That is to say, variations in scintillation parameters are the signature of the relative transverse motions of Earth, the pulsar, and the scattering medium. Periodic variations in $\tau_{\rm d}$ and $\nu_{\rm d}$, for instance, are caused by periodically changing transverse velocities of the pulsar and the Earth. Thus, long-term scintillation analyses with periodic variations provide us a possibility to investigate the IISM properties and orbital parameters \cite{rcn+14,rcb+20,mma+22}. 

Among the prominent features of the local ISM that could be investigated with pulsar scintillation, is the local (hot) bubble. This is a large void in the local Galactic neighbourhood likely created by past supernova explosions \cite{sc01,hei98}. Most likely the local bubble is filled with a hot ($\sim 10^6$K) ionised gas that is too tenuous to noticeably affect pulsar radiation, although the density of this gas is still a matter of debate \cite{ws09,gcc+14,scc+14}. It is very well documented, however, that the local bubble contains some interstellar clouds \cite{lr14,lrr08} and that it has a very well-defined boundary \cite{lvv+14} that can be studied through pulsar scintillation \cite{lrr08,bgr98}.

PSRs~J0613$-$0200 and J0636+5128 both are observed to have strong annual variations in the scintillation timescale, which was reported in Liu et al. (2022) \cite{lvm+22}. 
The main purpose of this work is to make use of these annual variations to investigate the IISM properties, including the position, the velocity and the anisotropy of the scattering region. Additionally, PSRs~J0613$-$0200 and J0636+5128 are both in a binary system. Another purpose in this work is to determine the orbital inclination angle $i$ and the longitude of ascending node $\Omega_{\rm asc}$ in the pulsar binary system because these two parameters are important for constraining neutron star masses \cite{dpr+10} but are difficult to measure through pulsar timing alone.  The low-mass companion star and short orbit of PSR~J0636+5128 only contribute a small amount to $V_{\rm{eff}}$, which leads to a negligible orbital period fluctuation in the scintillation parameters \cite{ksv+18}.  %Since only PSR~J0613$-$0200 is observed with a fluctuation caused by the changing pulsar orbital velocity, 
Therefore, we only investigate the orbital parameters for PSR~J0613$-$0200.

In this paper, we present the relevant scintillation background and construct our models assuming an anisotropic thin scattering screen in Section~\ref{sec:md}. In Section~\ref{sec:data}, we describe the pulsars and data information, and how we estimate the earth and pulsar velocities. We describe the Bayesian inference and the Markov chain Monte Carlo method that are used to fit the parameters in Section~\ref{sec:bayes}. The results and discussion are presented in Section~\ref{sec:results}. Our conclusions are in Section~\ref{sec:conclusion}.
%-----------------------------------------------------------------------------------------------------------------------------------------------
\section{Scintillation from a thin screen}
\label{sec:md}
Firstly, we assume that the scintillation is caused by a thin screen at distance $D_{s}$ from the Earth. For a thin screen model, Cordes \& Rickett (1998) \cite{cr98} presented that the scintillation velocity $V_{\rm ISS}$ can be estimated from the scintillation parameters,
\begin{equation}
\label{eqn:viss_1} 
V_{\rm ISS}=A_{\rm ISS}\frac{\sqrt{D\nu_{\rm d}}}{f\tau_{\rm d}},
\end{equation}
\noindent where $D$ is the pulsar distance in kpc, $f$ is the observing frequency in GHz, $\nu_{\rm d}$ is in units of MHz and $\tau_d$ is in units of seconds. 

The factor $A_{\rm ISS}$ depends on the assumed geometry of the scattering medium and the exponent $\alpha$ in the phase structure function on the scattering screen.
For a thin screen with Kolmogorov turbulence, $\alpha=5/3$, and $A_{\rm ISS} \approx 2.78\times10^{4} \sqrt{2D_s/(D-D_s})$, where $D_s$ is in units of kpc. However, this formulation is true for isotropic scattering, while scattering is often seen to be quite anisotropic \cite{cmr+05}.
Stinebring et al. (2022) \cite{srm+22} presented that about 20\% of the pulsars in their sample are observed with reverse arclets and a deep valley along the delay axis in the secondary spectrum, indicating substantial anisotropy of scattering. In this work, we consider an anisotropic scattering screen and anisotropy dependent $A_{\rm ISS}$. This anisotropy can be determined by the axial ratio $A_{\rm r}$ assuming the spatial diffraction pattern as an ellipse, where $A_{\rm r}$ = 1 indicates isotropic scattering. According to the physics of the phase structure function for anisotropic scattering outlined in Rickett et al. (2014) \cite{rcn+14}, taking a geometric mean of the spatial scales on the major and minor axes of anisotropic scattering, we consider that $A_{\rm ISS}$ is 2.78$\sqrt{(A_{r} + 1/A_{r})/2}\sqrt{2D_s/(D-D_s}) \times10^{4}$.

In the thin screen model, we can also predict the scintillation velocity $V_{\rm ISS}$ at the observer as 
\begin{equation}
\label{eqn:viss_2} 
V_{\rm ISS} = {|}V_{\rm eff}{|}\frac{D}{D-D_{s}}.
\end{equation}
Thus, Equation~\ref{eqn:viss_1} and Equation~\ref{eqn:viss_2} can be restructured as, 
\begin{equation}
\label{eqn:model} 
\frac{\sqrt{\nu_{\rm d}}}{f\tau_{\rm d}} \equiv Y = \frac{ {|}V_{\rm eff} {|}}{A_{\rm ISS}}\frac{\sqrt{D}}{D-D_{s}}
\end{equation}
\noindent where we combine all the scintillation observables in the definition of $Y$.

The effective velocity is a combination of the pulsar's, the Earth's and the IISM's transverse velocities weighted by the scattering screen distance as,
\begin{equation}
\label{eq:veff}
V_{\rm eff} =  \frac{D-D_s}{D}V_{\rm E} + \frac{D_s}{D}(V_{\rm p} + V_{\rm \mu}) - V_{\rm IISM},
\end{equation}
where $V_{\rm E}$ is the Earth velocity, $V_{\rm p}$ is the binary orbital transverse velocity, $V_{\rm \mu}$ is the pulsar proper motion transverse velocity and $V_{\rm IISM}$ is the transverse velocity of the IISM \cite{cr98}.

\subsection{Mildly anisotropic scattering}
In mildly anisotropic scattering, we take the effective transverse velocity in equatorial coordinates employing the quadratic form from Rickett et al. (2014) \cite{rcn+14},
\begin{equation}
\label{eq:viss_re} 
\begin{aligned}
&{|}V_{\rm eff}{|} = \sqrt{aV_{\rm eff, \alpha}^2+bV_{\rm eff, \delta}^2+cV_{\rm eff, \alpha}V_{\rm eff, \delta}},\\
&V_{\rm eff, \alpha} = \frac{D-D_s}{D}V_{\rm E, \alpha} + \frac{D_s}{D}(V_{\rm p, \alpha} + V_{\rm \mu, \alpha}) - V_{\rm IISM, \alpha},\\
&V_{\rm eff, \delta} = \frac{D-D_s}{D}V_{\rm E, \delta} + \frac{D_s}{D}(V_{\rm p, \delta} + V_{\rm \mu, \delta}) - V_{\rm IISM, \delta},\\
\end{aligned}
\end{equation}
where $V_{\rm eff, \alpha}$ and $V_{\rm eff, \delta}$ are the components of the effective velocity in right ascension ($\alpha$) and in declination ($\delta$), respectively. The coefficients a, b, and c are parameterized by the axial ratio $A_{\rm r}$ and the orientation angle of the major axis $\psi$ considering the spatial diffraction pattern 
as an ellipse,
\begin{equation}
\label{eqn:abc}
\begin{aligned}
a &=  \left[1 - R\cos{\left(2\psi\right)}\right]/\sqrt{1 - R^2}, \\ 
b &=  \left[1 + R\cos{\left(2\psi\right)}\right]/\sqrt{1 - R^2}, \\ 
c &=  - 2R\sin{\left(2\psi\right)}/\sqrt{1 - R^2},
\end{aligned}
\end{equation}
where the orientation angle $\psi$ is clockwise from the right ascension of the pulsar and $R$ is a scaled parameter for $A_{\rm r}$: $R=(A_{\rm r}^2 - 1)/(A_{\rm r}^2 + 1)$, which is bound between 0 (indicating that the inhomogeneities in the plasma density are isotropic) and 1 (indicating that the inhomogeneities in the plasma density are extremely anisotropic). 

\subsection{Extremely anisotropic scattering}
Since Brisken et al. (2010) \cite{bmg+10} measured an extremely anisotropic distribution of images for PSR~B0834+06 at 327\,MHz using very long baseline interferometry, 1D scattering screen models are assumed in increasing numbers of studies\cite{msa+20,mbm+22}. In the 1D scattering screen model, the effective transverse velocity is only sensitive to the major axis on the sky, 
\begin{equation}
\label{eq:viss_re1} 
\begin{aligned}
&{|}V_{\rm eff}{|} = {\Big{|}}\sqrt{aV_{\rm eff, \alpha}^2+bV_{\rm eff, \delta}^2+cV_{\rm eff, \alpha}V_{\rm eff, \delta}} - V_{\rm IISM}{\Big{|}},\\
&V_{\rm eff, \alpha} = \frac{D-D_s}{D}V_{\rm E, \alpha} + \frac{D_s}{D}(V_{\rm p, \alpha} + V_{\rm \mu, \alpha}),\\
&V_{\rm eff, \delta} = \frac{D-D_s}{D}V_{\rm E, \delta} + \frac{D_s}{D}(V_{\rm p, \delta} + V_{\rm \mu, \delta}),\\
\end{aligned}
\end{equation}
where $V_{\rm IISM}$ is the IISM velocity on the major axis, the coefficients a, b, and c are re-written as,
\begin{equation}
\label{eqn:abc_1D}
\begin{aligned}
a &=  \sin^2{\psi}, \\ 
b &=  \cos^2{\psi}, \\ 
c &=  - 2\sin{\psi}\cos{\psi}.
\end{aligned}
\end{equation}

Extremely anisotropic scattering indicates a large $A_{\rm r}$, in this case, the factor $A_{\rm ISS}$ cannot be inferred by the functions of Rickett et al. (2014) \cite{rcn+14}. We therefore give a free scaling index $K$ (instead of a fixed constant in the isotropic scattering model) to $A_{\rm ISS}$, thus, $A_{\rm ISS} = K\sqrt{2D_s/(D-D_s}) \times10^{4}$.

\section{Pulsar and data information}
\label{sec:data}

\begin{figure*}
\centering
\includegraphics[scale=0.5]{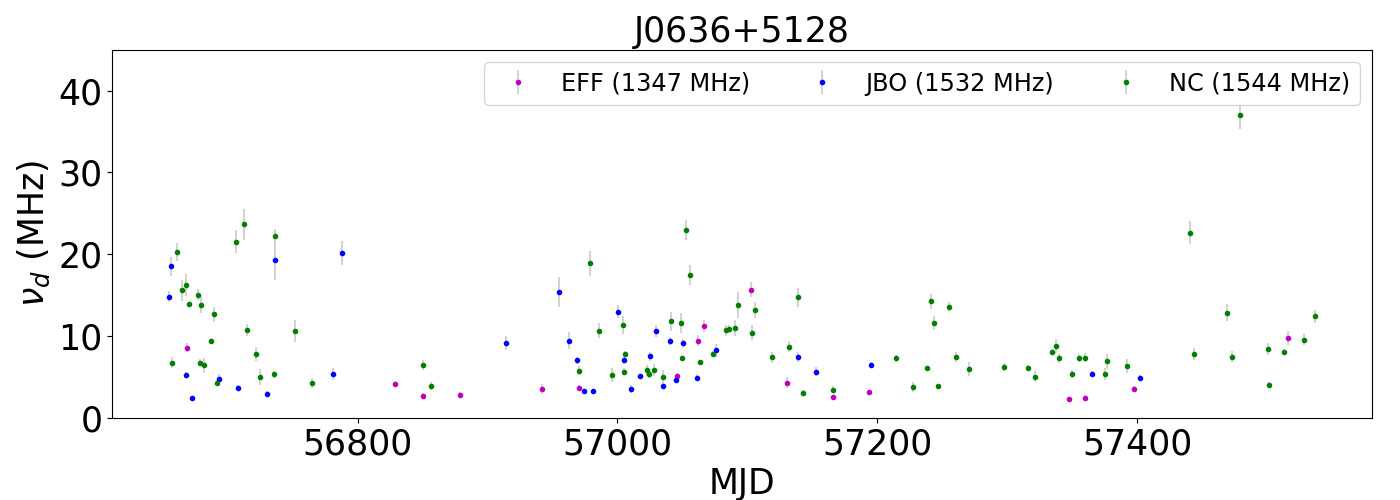}
\caption{Measurements of $\nu_{\rm d}$ as a function of time for PSR J0636+5128. These measurements are based on the same data presented by Liu et al. (2022) \cite{lvm+22} but have been revised in order to correct for skewness in the dynamic spectrum's ACF, caused by a transverse phase gradient in the scattering screen. Only observations until MJD 57600 are used in the present work since the observing cadence significantly decreased after that (see Liu et al. (2022) \cite{lvm+22}).}
\label{fig:scbw0636}
\end{figure*}

\begin{table*}
\footnotesize
\begin{threeparttable}
\caption{Measured scintillation parameters and observation information}
\label{tab:obsinfo} 
\doublerulesep 0.1pt 
\tabcolsep 10pt %space between two columns. 
\begin{tabular}{cccccccccccc}
\toprule
Pulsar  & Telescope & $f_{\rm c}$ & MJD range & N$_{\rm{obs}}$ & $\Delta{f_{\rm ch}}$ & BW & t$_{\rm{sub}}$ & $\bar{\rm{t}}_{\rm{obs}}$ & $\nu_{\rm d}$&$\tau_{\rm d}$\\
	&&{\small(MHz)}&&&{\small(MHz)}&{\small(MHz)}&{\small(sec)}&{\small(min)}&{\small(MHz)}&{\small(min)}\\
\hline
J0613$-$0200&EFF&1349&55661--58019&55&1.56&140.6&10&34&$\textcolor{red}{1.7}^{3.3}_{\textcolor{red}{1.1}}$&$11^{24}_{7}$\\
$\cdots$&JBO&1556&56167--58857&89&1&352.0&65&35&$2.3^{4.2}_{\textcolor{red}{1.3}}$&$13^{\textcolor{red}{27}}_{8}$\\
$\cdots$&NRT&1484&55817--58852&194&4&512.0&61&50&$\textcolor{red}{4.2}^{9.0}_{\textcolor{red}{3.0}}$&$16^{40}_{9}$\\
\\
J0636+5128&EFF&1347&56669--59021&23&1.56&200.0&10&28&$6.4^{37.5}_{3.4}$&$8.8^{\textcolor{red}{48.8}}_{4.6}$\\
$\cdots$&JBO&1532&56655--58120&35&2&384.0&60&29&$9.7^{17.1}_{4.8}$&$8.0^{\textcolor{red}{32.6}}_{5.1}$\\
$\cdots$&NRT&1544&56657--58419&97&4&392.0&61&60&$12^{40}_{\textcolor{red}{5}}$&$11^{\textcolor{red}{56}}_{5}$\\
\bottomrule
\end{tabular}
\begin{tablenotes}
\item[1)] $f_c$ is the effective center frequency, N$_{\rm{obs}}$ is the number of observations, $\Delta{f_{\rm ch}}$ is the channel bandwidth, BW is the effective bandwidth of the observation after the removal of RFI, t$_{\rm{sub}}$ is the subintegration length, $\bar{\rm{t}}_{\rm{obs}}$ is the mean observation length.  $\nu_{\rm d}$ and $\tau_{\rm d}$ are the scintillation bandwidth and timescale, respectively. Values in red indicate that measurements are to be considered as upper (for $\nu_{\rm d}$) or lower (for $\tau_{\rm d}$) limits in Liu et al. (2022) \cite{lvm+22}.
\end{tablenotes}
\end{threeparttable}
\end{table*}

\begin{table*}
\footnotesize
\begin{threeparttable}
 \caption{Parameters of MSPs~J0613$-$0200 and J0636+5128}
  \label{tab:pulsars_info}  
\doublerulesep 0.2pt 
\tabcolsep 14pt %space between two columns. 
\begin{tabular}{lll}
\toprule
  \multicolumn{3}{c}{PSR~J0613-0200} \\
  \\
 
Parameter Name & Parameter value & Reference paper\\
Right ascension, $\alpha$ (J2000)  & 06:13:43.975672& Desvignes et al. (2016) \cite{dcl+16}\\
Declination, $\delta$ (J2000)  & $-$02:00:47.22533 &Desvignes et al. (2016) \cite{dcl+16}\\
Reference epoch for $\alpha$ and $\delta$,& 55000.0& Desvignes et al. (2016) \cite{dcl+16}\\
Proper motion in $\alpha$, $\mu_{\rm \alpha}$ (mas yr$^{-1}$)   & 1.822(8) & Desvignes et al. (2016) \cite{dcl+16}\\ 
Proper motion in $\delta$, $\mu_{\rm \delta}$ (mas yr$^{-1}$)   & $-$10.36(2) & Desvignes et al. (2016) \cite{dcl+16}\\
Pulsar distance, $D$ (kpc)  & 1.1(2) & Matthews et al. (2016) \cite{mnf+16}\\
Orbital period, $P_{\rm b}$ (days) & 1.198512575184& Desvignes et al. (2016) \cite{dcl+16}\\
Epoch of periastron passage, $T_{0}$ (MJD)   &53113.953 & Desvignes et al. (2016) \cite{dcl+16}\\
Projected semi-major axis, $x$ (s)    & 1.09144409 & Desvignes et al. (2016) \cite{dcl+16}\\
Longitude of periastron, $\omega$ ($^{\circ}$) & 47.2 & Desvignes et al. (2016) \cite{dcl+16}\\
Orbital eccentricity, $e_{\rm b}$ (10$^{-6}$)   & 5.4& Desvignes et al. (2016) \cite{dcl+16}\\
Orbital inclination, $i$ ($^{\circ}$)    & 66$^{+7}_{-10}$& Fonseca et al. (2016) \cite{fpe+16}\\
Sine of inclination, $\sin(i)$ &                  0.94(2) & Perera et al. (2019)  \cite{pdd+19}\\
%Longitude of the ascending node,  $\Omega_{\rm asc}$ & $\ast$  &  $\ast$ \\
 \\
 \multicolumn{3}{c}{PSR~J0636+5128} \\
 \\
 
 Parameter Name & Parameter value & Reference paper\\
Right ascension, $\alpha$ (J2000)  &06:36:04.84645 & Stovall et al. (2014) \cite{slr+14}\\
Declination, $\delta$ (J2000)  & 51:28:59.9625 & Stovall et al. (2014) \cite{slr+14}\\
Reference epoch for $\alpha$ and $\delta$ & 56307.0& Stovall et al. (2014) \cite{slr+14}\\
Proper motion in $\alpha$, $\mu_{\rm \alpha}$ (mas yr$^{-1}$)   & 4.3(9) & Stovall et al. (2014)\cite{slr+14}\\ 
Proper motion in $\delta$, $\mu_{\rm \delta}$ (mas yr$^{-1}$)   & 2(1) & Stovall et al. (2014) \cite{slr+14}\\
Pulsar distance, $D$ (kpc)  & 1.1$\pm$0.25 & Kaplan et al. (2018) \cite{ksv+18}\\

Orbital period, $P_{\rm b}$ (days) & 0.066551340763&  Alam et al. (2021) \cite{aab+21}    \\
%Epoch of periastron passage, $T_{0}$ (MJD)   &  $\ast$ & $\ast$\\
Projected semi-major axis, $x$ (s)    & 0.00898636 & Alam et al. (2021) \cite{aab+21}\\
Longitude of periastron, $\omega$ ($^{\circ}$) & 5 &  Alam et al. (2021) \cite{aab+21}\\
Orbital eccentricity, $e_{\rm b}$ (10$^{-5}$)   & 1.7& Alam et al. (2021) \cite{aab+21}\\
Orbital inclination, $i$ ($^{\circ}$)    & 24.3$\pm$3.5 & Kaplan et al. (2018) \cite{ksv+18}\\
%Longitude of the ascending node,  $\Omega_{\rm asc}$ &  $\ast$ &  $\ast$ \\

\bottomrule
\end{tabular}
\end{threeparttable}
\end{table*}

Equation~\ref{eqn:model} is the embodiment of our model, where the left side of the equation can be seen as the measurement of $Y$, and the right side is the prediction of $Y$.
This work is a deeper analysis of Liu et al. (2022) \cite{lvm+22} that presented the annual variations of scintillation parameters of PSRs~J0613$-$0200 and J0636+5128. The scintillation parameters ($\nu_{\rm d}$ and $\tau_{\rm d}$) in that literature are estimated by exploiting the auto-correlation function (ACF) of dynamic spectra: $\nu_{\rm d}$ is the half-width at a half maximum along the frequency axis and $\tau_{\rm d}$ is the half-width at 1/e along the time axis in the two-dimensional ACF. All observations are from three telescopes: the Effelsberg 100-m Radio Telescope (EFF), the Lovell Radio Telescope at the Jodrell Bank Observatory (JBO) and the Nan{\c{c}}ay radio telescope (NRT). The median and the 5/95 percentiles of the measured scintillation parameters and more observation information are listed in Table~\ref{tab:obsinfo}. More details about the data analysis are described Liu et al. (2022) \cite{lvm+22}.

Due to the limited frequency resolution, most observations of PSR~J0613$-$0200 from NRT and EFF only give upper limits for $\nu_{\rm d}$. After several experiments, we noticed that the contribution from the variation of $\nu_{\rm d}$ to the fitted parameters is negligible, whereas the contribution from the average value is large. We therefore use the average value of the $\nu_{\rm d}$ of JBO observations, which have the highest frequency resolution and have reliable measures of $\nu_{\rm d}$, instead of the instantaneous values for PSR~J0613$-$0200. As the ACF of some observations of PSR~J0636+5128 shows a skewness resulting from a transverse phase gradient of the scattering screen, the scintillation bandwidths are somewhat biased \cite{yzm+21}. We use Equation~A6 of Rickett et al. (2014) \cite{rcn+14} to correct the skewness for such ACFs, selecting the best fit by eye, and then re-evaluate $\nu_{\rm d}$. The revised values of $\nu_{\rm d}$ are plotted in Figure~\ref{fig:scbw0636}. Additionally, as the strength of scintillation for PSR~J0636+5128 changed significantly and the sampling density drops sharply after MJD~57600, we only use the scintillation measurements before MJD~57600.

We estimate the uncertainty of $Y$ using the fitting procedure and the statistical uncertainties of scintillation parameters. As described in Liu et al. (2022) \cite{lvm+22}, the uncertainty derived from the ACF is heavily underestimated, we thus include the noise parameters EFAC (an additional scaling factor) and EQUAD (an additional variance added in quadrature) on errors of $Y$ in the following MCMC fitting. In the following context, we use factors $F$ and $Q$ to denote the noise parameters EFAC and EQUAD, respectively.

On the right side of Equation~\ref{eqn:model}, $V_{\rm eff}$ contains four terms of transverse velocity: the pulsar's proper motion, the pulsar's binary motion, the Earth's orbital motion, and the velocity of the scattering screen.
To obtain Earth's velocity, we use a Cartesian coordinate system with the Solar System barycenter as the origin employing the CALCEPH software package \cite{glm+15}, then project it onto equatorial coordinates.
The pulsar transverse velocity can be estimated using the proper motion $\mu$ that was measured with high precision using timing, $V_{\rm \mu} = 4.74 \mu D$. Assuming a non-relativistic binary system, we need five Keplerian parameters: $P_{b}$, $T_{0}$, $x$, $\omega$ and $e$ and two additional parameters: $i$ and $\Omega_{\rm asc}$ to estimate the pulsar orbital velocity. These five Keplerian parameters are usually well measured with timing. Firstly, the pulsar mean orbital velocity $V_{0}$ is given by $V_0 = 2\pi {x}c/(\sin\,iP_{\rm b}\,(1-e^2 )^{1/2} )$, $x$ is the projected semi-major axis in units of seconds, $c$ is the speed of light in vacuum \cite{lk+04}. The orbital velocity is projected along and perpendicular to the line of nodes in the plane of the sky, marked as $V_{\rm p,\parallel}$ and $V_{\rm p,\perp}$,
\begin{equation}
\begin{aligned}
\label{eq:orbittwo}
&V_{\rm p,\parallel} = -V_0\,(e\sin{\omega}+\sin{\phi}) \\
&V_{\rm p,\perp} = V_0\cos{i}\,(e{\cos}\omega+\cos{\phi}) \\
\end{aligned}
\end{equation}
where $\phi$ is the orbital phase from the line of nodes, $\phi = \theta + \omega$, $\theta$ is the true anomaly and can be calculated by $P_{\rm b}$, $T_{0}$ and $e$.
Then, the transverse orbital velocity in the directions of right ascension and declination are 
\begin{equation}
\begin{pmatrix} V_{\rm p, \rm \alpha} \\ V_{\rm p, \rm \delta}  \end{pmatrix} = \begin{pmatrix} \cos{\Omega_{\rm asc}} & \sin{\Omega_{\rm asc}} \\ -\sin{\Omega_{\rm asc}} &\cos{\Omega_{\rm asc}}  \end{pmatrix}\begin{pmatrix} V_{\rm p,\parallel} \\ V_{\rm p,\perp}  \end{pmatrix}.
\end{equation}

In the next two subsections, we introduce the two pulsars we study in this work, and describe their parameters.

%-------------------------------------------------------

\subsection{PSR~J0613$-$0200}
\label{sec:j0613}
PSR~J0613$-$0200 is a binary millisecond pulsar (MSP) discovered by Lorimer (1995)  \cite{lor95}. This pulsar is included in all current ongoing PTA experiments \cite{vlh+16}.
Main et al. (2020) \cite{msa+20} observed scintillation arcs in the 2D power spectrum of scintillation using data from the Large European Array for Pulsars and the Effelsberg telescope, finding the arc curvature varies annually (although they ignored the orbital contribution).

All known parameters used in this work have been listed in Table~\ref{tab:pulsars_info}, including the position, the proper motion and five Keplerian parameters \cite{dcl+16}, and the pulsar distance \cite{mnf+16}. Fonseca et al. (2016) \cite{fpe+16} and Perera et al. (2019) \cite{pdd+19} determined this pulsar's binary orbital inclination angle $i$ using timing.
Since only the value of $\sin(i)$ can be determined using timing, we investigate the sense of inclination (i.e., whether $i > 90^{\circ}$ or $i<90^{\circ}$) in this work.
In addition, there has been no published measurement of the longitude of ascending node $\Omega_{\rm asc}$, which we also investigate in this work.
The position and the five Keplerian parameters of the binary system for this pulsar have been determined with high precision, and we therefore fix them in the estimation of $V_{\rm eff}$. In contrast, the uncertainties of the distance and proper motion are included as Gaussian priors. 

In conclusion, the scintillation velocity model of PSR~J0613$-$0200 includes nine free parameters in the mildly anisotropic scattering case: $D_{s}$, $V_{\rm IISM, \alpha}$, $V_{\rm IISM, \delta}$, $A_{R}$, $\psi$, $i$, $\Omega_{\rm asc}$, $F$, $Q$; eight free parameters in the extremely anisotropic scattering case: $D_{s}$, $V_{\rm IISM}$, $K$, $\psi$, $i$, $\Omega_{\rm asc}$, $F$, $Q$.
%---------------------------------------------

\subsection{PSR~J0636+5128}
\label{sec:j0636}
PSR~J0636+5128 is also a binary MSP, discovered in the Green Bank Northern Celestial Cap Pulsar Survey \cite{slr+14}. 
The position, the proper motion and the distance of this pulsar come from Stovall et al. (2014) \cite{slr+14}, with details shown in Table~\ref{tab:pulsars_info}. Similar to PSR~J0613$-$0200, the uncertainties of the pulsar distance and proper motion are included in the estimation of $V_{\rm eff}$.
For this pulsar, the orbit period is comparable to the scintillation timescale, and the companion is very light. Therefore the pulsar orbital velocity is low, which means that scintillation parameters do not display measurable fluctuations from the orbital velocity, and the orbital parameter $T_{0}$ has not yet been published.
Thus, we ignore the pulsar orbital velocity in the estimation of $V_{\rm eff}$.
Finally, the scintillation velocity model of PSR~J0636+5128 includes seven free parameters in the mildly anisotropic scattering case: $D_{s}$, $V_{\rm IISM, \alpha}$, $V_{\rm IISM, \delta}$, $A_{r}$, $\psi$, $F$, $Q$; six free parameters in the extremely anisotropic scattering case: $D_{s}$, $V_{\rm IISM}$, $K$, $\psi$, $F$, $Q$.
%------------------------------------------------------------------------------------------------------------------------

\section{Bayesian inference and the Markov chain Monte Carlo method}
\label{sec:bayes}

We use Bayesian theory to derive the parameters of the investigation list because it is capable of performing a robust mean for parameter estimation, model selection and visualization of parameter correlations \cite{ahl+19,rch+19}.
From Bayes theorem we have the joint posterior for a set of model parameters $\theta$,
\begin{equation}
\label{eqn:posterior} 
p(\theta|d)=\frac{p(d|\theta)p(\theta)}{Z}
\end{equation}
where $p(\theta)$ is the prior probability distribution for parameters, $p(d|\theta)$ is the likelihood function of the data $d$ given a set of parameters of the model and $Z$ is the fully marginalized likelihood function (alternatively the evidence, the prior predictive and the marginal density of the data). $Z$ is a normalising constant for the posterior and is often ignored in a single model.

To calculate numerical approximations of multi-dimensional integrals, the Markov chain Monte Carlo (MCMC) method was proposed in the last century. The MCMC method comprises a class of algorithms for sampling from high-dimensional probability distributions. Markov Chain is a mathematical system that experiences transitions from one state to another according to certain probabilistic rules. Hastings (1970) \cite{has70} presented Monte Carlo sampling methods using Markov Chains and their applications. In this work, we use a stable and well-tested Python implementation {\bf emcee} \cite{fhl+13} of the affine invariant ensemble sampler \cite{gw+10} for the MCMC method. 
Employing Bayesian theory and {\bf emcee}, we can produce and draw posterior probability distributions for all studied parameters.

We give a uniform prior within a physically-motivated bound region for each parameter: $D_{s}\in (0,D)$, $V_{\rm IISM, \alpha}$ and $V_{\rm IISM, \delta} \in [-100,100]$\,km\,s$^{-1}$, $A_{r} \in [1,100]$, $\psi \in [0^{\circ},180^{\circ})$, $i \in [0^{\circ},180^{\circ})$, $\Omega_{\rm asc} \in [-180^{\circ},180^{\circ})$, and run {\bf emcee} with 10$^5$ steps. The first 25\% of the steps in the sampling process are considered to be 'burn-in' stage and are therefore discarded. In that stage, the chain is not effectively exploring the parameter space as the global maximum is not yet found.

%--------------------------------------------------------------------
%------------------------------------------------------------------------------------------------------------------------

\section{Results and discussion}
\begin{figure*}
\centering
\includegraphics[scale=0.33]{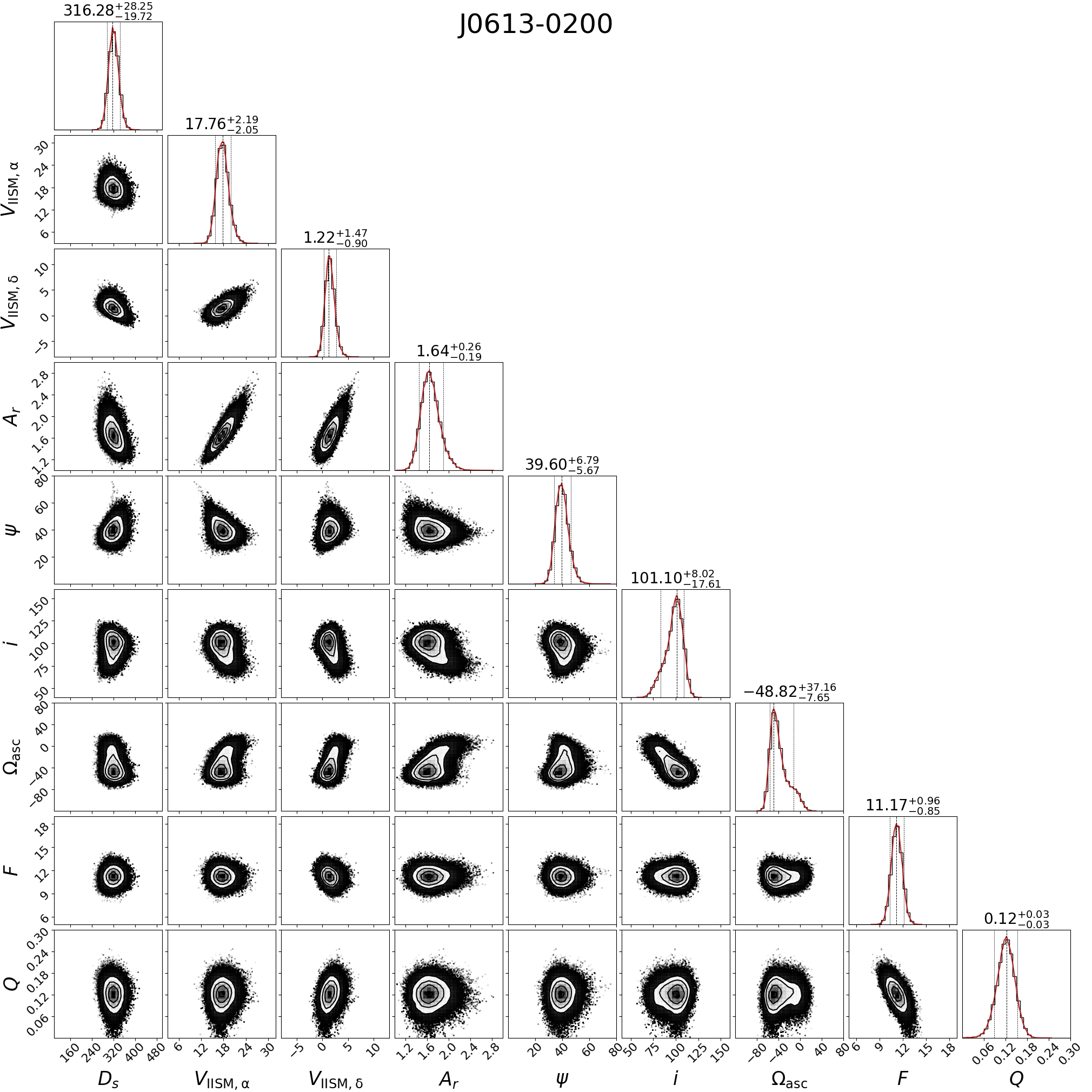}
\caption{The posterior probability distributions of all fitted parameters for PSR~J0613$-$0200. In the 1-D histograms, the red lines are the kernel-density estimate smoothed versions of the distributions, the vertical dashed lines indicate the 10\% fractional percentiles, the most likely values and the 90\% fractional percentiles. The most likely values and the upper/lower errors are presented at the top of the 1-D histograms. Factors $F$ and $Q$ are the noise parameters EFAC and EQUAD, respectively. The factor $Q$ is scaled up by a factor of one thousand, and $D_{s}$ is in units of pc.}
\label{fig:mcmc_0613}
\end{figure*}

\begin{figure*}
\centering
\includegraphics[scale=0.4]{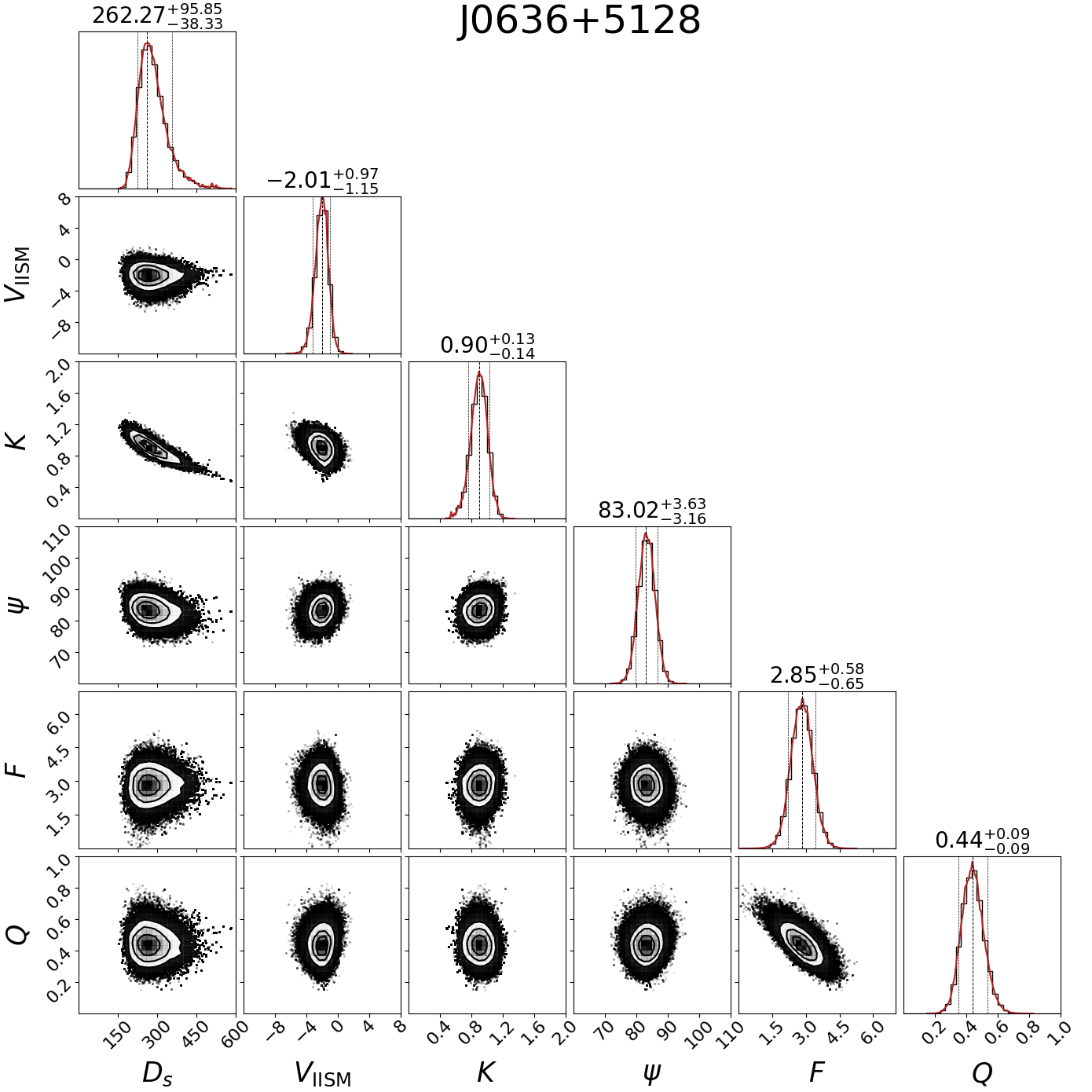}
\caption{The posterior probability distributions of all fitted parameters for PSR~J0636+5128. In the 1-D histograms, the red lines are the kernel-density estimate smoothed versions of the distributions, the vertical dashed lines indicate the 10\% fractional percentiles, the most likely values and the 90\% fractional percentiles. The most likely values and the upper/lower errors are presented at the top of the 1-D histograms. Factors $F$ and $Q$ are the noise parameters EFAC and EQUAD, respectively. The factor $Q$ is scaled up by a factor of one thousand, and $D_{s}$ is in units of pc.}
\label{fig:mcmc_0636}
\end{figure*}

We show the best-fit results for PSRs~J0613$-$0200 and J0636+5128 in Figures~\ref{fig:mcmc_0613} and \ref{fig:mcmc_0636}, respectively, including the most likely values with uncertainties (the 1-D and 2-D posterior probability distributions) for all fitted parameters. 

\begin{figure}[H]
\centering
\includegraphics[scale=0.26]{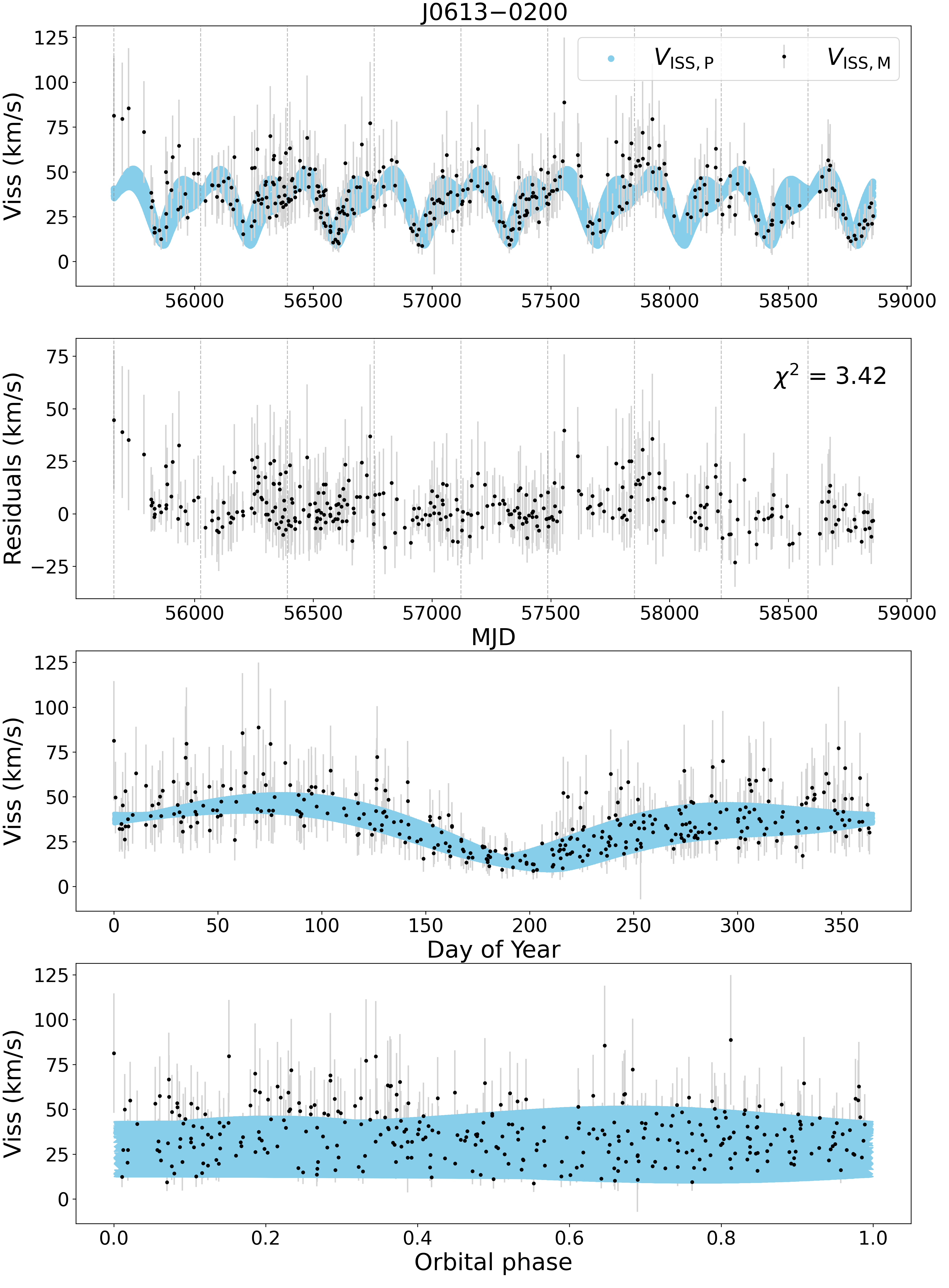}
\caption{The scintillation velocities of PSR~J0613$-$0200. $V_{\rm{ISS, M}}$ is the scintillation velocity from Equation~\ref{eqn:viss_1}, $V_{\rm{ISS, P}}$ is that from Equation~\ref{eqn:viss_2}. The $V_{\rm{ISS, P}}$ values in the first panel are determined using a cadence of 0.1\,days.  The second panel shows the differences between $V_{\rm{ISS, M}}$ and $V_{\rm{ISS, P}}$. In the third and fourth panels, we plot all points as a function of day of year and orbital phase, respectively.}
\label{fig:variation0613}
\end{figure}

\begin{figure}[H]
\centering
\includegraphics[scale=0.26]{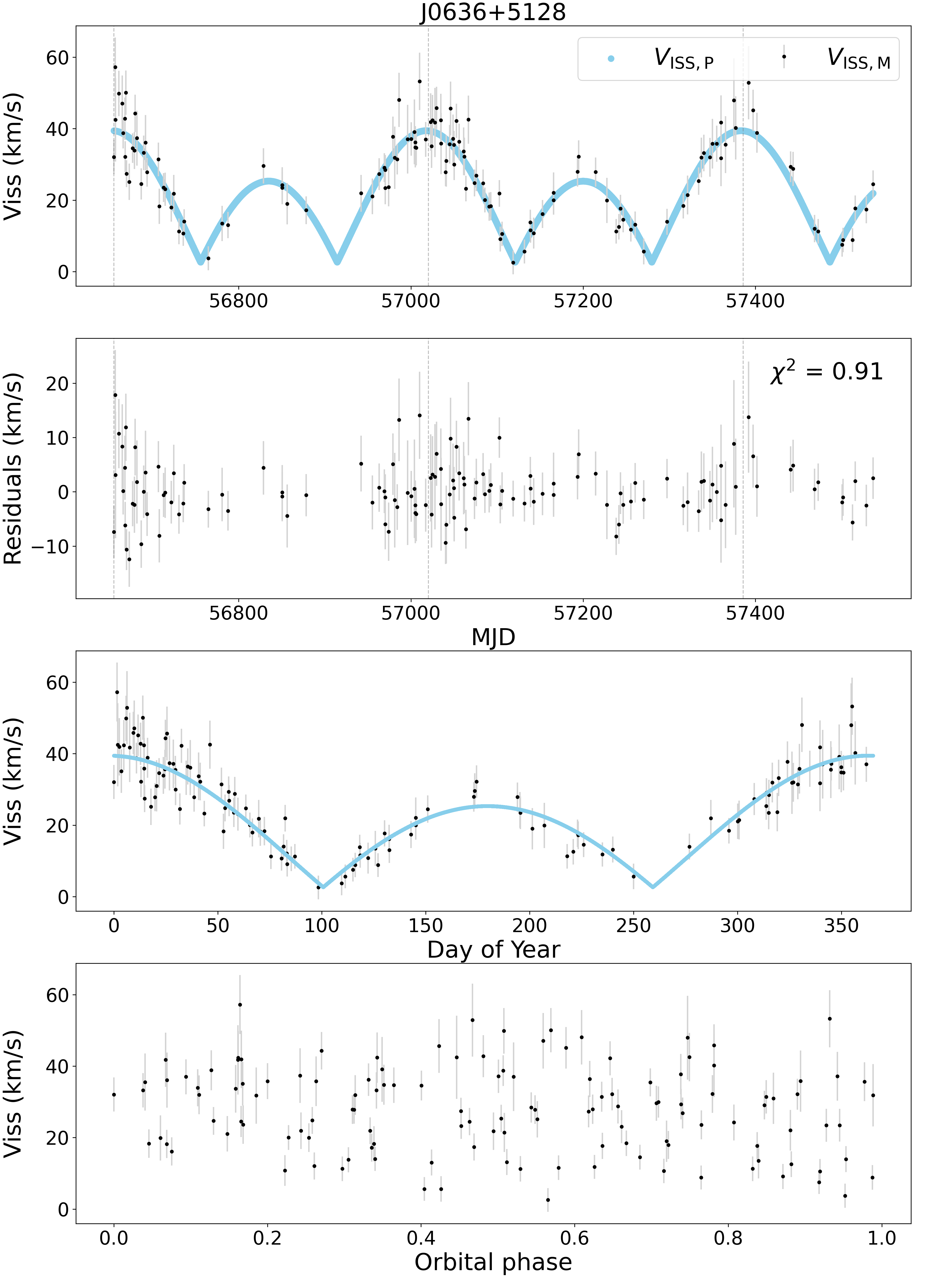}
\caption{The scintillation velocities of PSR~J0636+5128.  $V_{\rm{ISS, M}}$ is the scintillation velocity from Equation~\ref{eqn:viss_1}, $V_{\rm{ISS, P}}$
 is that from Equation~\ref{eqn:viss_2}. The $V_{\rm{ISS, P}}$ values in the first panel are determined using a cadence of 0.1\,days.  The second panel shows the differences between $V_{\rm{ISS, M}}$ and $V_{\rm{ISS, P}}$. In the third and fourth panels, we plot all points as a function of day of year and orbital phase, respectively. Since the absence of the epoch of periastron passage $T_{0}$, $V_{\rm ISS, P}$ is omitted from the fourth panel.}
\label{fig:variation0636}
\end{figure}

We use Equation~\ref{eqn:model} in the MCMC fitting to separate the scintillation observables from our scintillation velocity model. In
order to show the results from the MCMC fitting in the physical sense, we plot the scintillation velocities from Equations~\ref{eqn:viss_1} and
\ref{eqn:viss_2} for PSRs~J0613$-$0200 and J0636+5128 in Figures~\ref{fig:variation0613} and \ref{fig:variation0636}, respectively. The $V_{\rm{ISS, M}}$ is the scintillation velocity from Equation~\ref{eqn:viss_1}, $V_{\rm{ISS, P}}$ is that from Equation~\ref{eqn:viss_2}. We plot the differences between the two scintillation velocities and present the reduced $\chi^{2}$ in the second panel. In the third and fourth panels, all
scintillation velocities are plotted as a function of the day of year and orbital phase, respectively.

\label{sec:results}

For PSR~J0613$-$0200, the best fit is from the mildly anisotropic scattering model. In Figure~\ref{fig:mcmc_0613}, all fitted parameters converge rapidly and most parameters show a well-behaved Gaussian distribution in the posterior probability distribution. The reduced $\chi^{2}$ in the mildly anisotropic scattering model is 3.42 which is better than that of the extremely anisotropic scattering model (6.52).
The best fit for PSR~J0636+5128 is from the extremely anisotropic scattering model and is shown in Figure~\ref{fig:mcmc_0636},  where the reduced $\chi^{2}$ of the scintillation velocities is 0.91. Also, all fitted parameters for this pulsar converge rapidly and show a well-behaved Gaussian distribution.

\subsection{Positions and velocities of the scintillation screens}

\begin{figure*}
\centering
\includegraphics[scale=0.45]{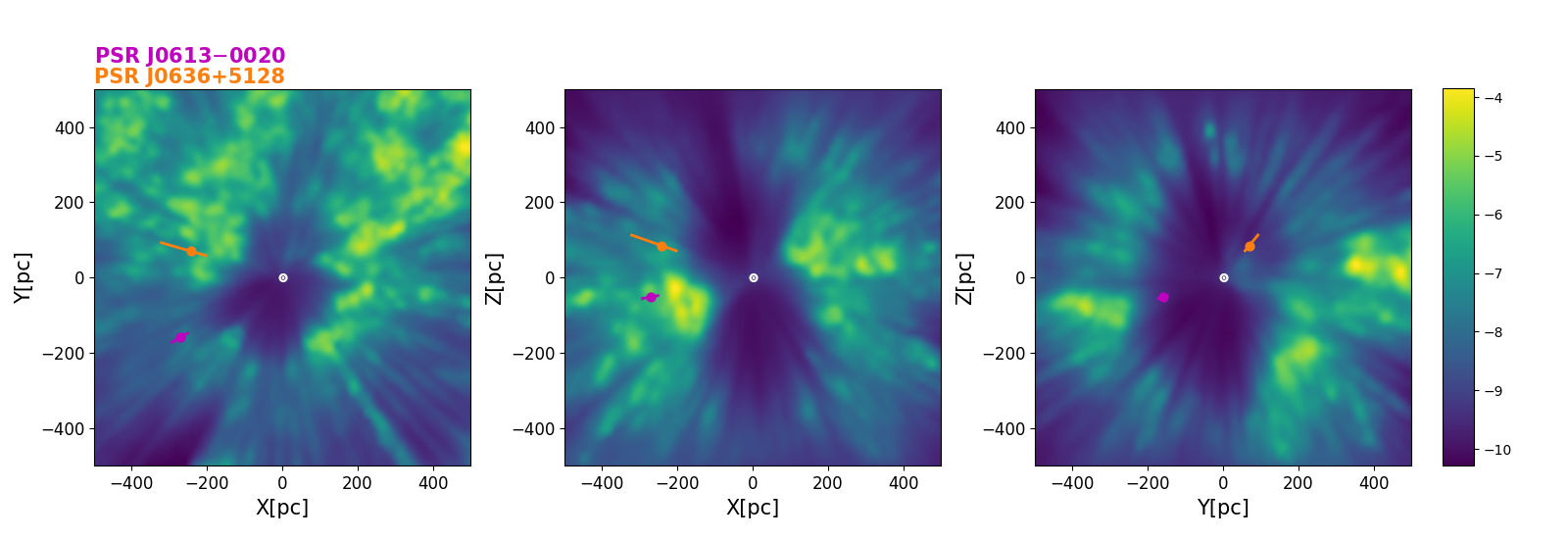}
\caption{Scintillation screen position on three crosscuts (XY, XZ and YZ planes) of the 3D dust extinction map from  Lallement et al. 2019 \cite{lbv+19}. The X axis points from the Sun (the circle dot) to the Galactic center, the Y axis points towards $l = 90^{\circ}$ and the Z axis points to the North Galactic pole at Galactic latitude $b = 90^{\circ}$. The colour scale shows $\log(A^{\prime}_{\rm v})$ and indicates the gas density, where $A^{\prime}_{\rm v}$ is the differential extinction in units of magnitudes per parsec.}
\label{fig:LB_screen}
\end{figure*}

\begin{figure*}
\centering
\includegraphics[scale=0.55]{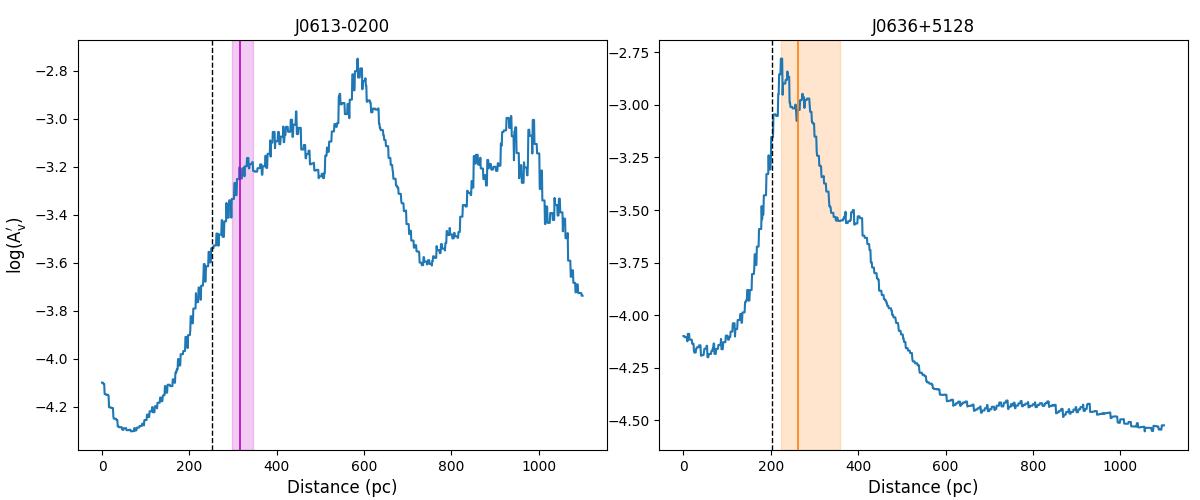}
\caption{The gas density distribution along the line of sight to two pulsars, from the 3D dust extinction map of Lallement et al. 2019 \cite{lbv+19}. The coloured vertical lines denote the estimated position of the scattering screen for each pulsar, while the lighter shades of the vertical lines indicate the uncertainty range of the scintillation screen positions. The dashed black vertical lines indicate positions of the inner local bubble surface on the line of sight to each pulsar.}
\label{fig:los}
\end{figure*}

Based on the posterior probability distributions discussed above, the most likely scintillation screen distances for PSRs~J0613$-$0200 and J0636+5128 are, respectively, $316^{+28}_{-20}\,$pc\footnote{We note our scintillation screen distance to PSR~J0613$-$0200 is consistent with  the one derived by Main et al. (2020) \cite{msa+20} and Main et al. (2023) \cite{mac+23}} and $262^{+96}_{-38}\,$pc. Both of these values are much closer to the Earth than to the pulsar; and they are in the vicinity of the Local Bubble boundary. In this section we will investigate this possible association.

In Figure~\ref{fig:LB_screen}, we plot three crosscuts of the solar neighborhood based on the 3D dust extinction map of Lallement et al. (2019) \cite{lbv+19}, which can be assumed a proxy for the gas density map. The three crosscuts clearly show the Local Bubble cavity. We overlay the projections of the scintillation screen positions (with error bars) for our two pulsars on the three crosscuts, which show that the scintillation screens of PSRs~J0613$-$0200 and PSR~J0636+5128 align well with the Local Bubble boundary.

Next, we calculate the distances of the inner local bubble surface on the line{s} of sight to our two pulsars using the Local Bubble inner surface model from Pelgrims et al. (2020) \cite{pfb+20}, finding a surface distance of 252\,pc towards PSR~J0613$-$0200 and 202\,pc towards PSR~J0636+5128. Both measured screen distances are slightly beyond this boundary, indicating the scattering screens of our pulsars are likely associated with the Local Bubble shell.

In Figure~\ref{fig:los}, we plot the gas density distribution along the line of sight of two pulsars employing the 3D dust extinction map of Lallement et al.\ 2019 \cite{lbv+19}, and mark the positions of the scattering screens and the inner local bubble surfaces. The positions of the scattering screens coincide with the first peaks of the gas density distributions along the lines of sight to PSRs~J0613$-$0200 and J0636+5128, respectively, which supports the hypothesis that the nearest dense IISM structure is the primary cause of the observed interstellar scintillation, assuming a positive correlation between the neutral and ionized parts of the medium.
Using this hypothesis, we estimate the lower limit of the shell thickness for the local bubble to be  twice the distance between the scattering screen and the inner local bubble surface. In the directions of PSRs~J0613$-$0200 and J0636+5128, the lower limits on the shell thickness are 128$^{+28}_{-20}$\,pc and 120$^{+96}_{-38}$\,pc, respectively.

The transverse IISM velocities for PSRs~J0613$-$0200 and J0636+5128 are around 18\,km/s and $-$2\,km/s, respectively. These velocities are well within the range of expectations for the ISM residing on the Local Bubble shell. For PSR~J0613$-$0200, the projection of the transverse IISM velocities onto the major axis of the IISM is roughly 12 km/s, which is in agreement with that of the 2013 event reported in Main et al. (2020) \cite{msa+20}. Main et al. (2023) \cite{mac+23} presented the IISM velocities towards PSR~J0613$-$0200 in both directions of right ascension and declination. In our estimates, the IISM velocity towards PSR~J0613$-$0200 in right ascension is 17.76$^{+2.19}_{-2.05}$\,km/s, which is consistent with their result. However, in the direction of declination,  the IISM velocity in our estimates is 1.22$^{+1.47}_{-0.90}$\,km/s, which slightly deviates from their result.

\subsection{Anisotropic scattering}
From our modelling, the scattering screens are both evidently anisotropic.
The axial ratio $A_{\rm r}$ of PSR~J0613$-$0200 is 1.6$^{+0.3}_{-0.2}$, which is in agreement with the mildly anisotropic scattering model. The orientation of the major axis $\psi$ is $40^{+7}_{-6}$ degrees, which is consistent with the orientation reported for the 2013 event in Main et al. (2020) \cite{msa+20}, but differs from the orientations reported in Main et al. (2023)\footnote{Main et al. (2023) \cite{mac+23} defined that $\psi$ is clockwise from the declination of the pulsar.}\cite{mac+23}.
For PSR~J0636+5128, the periodic variation of scintillation parameters is most accurately explained by the extremely anisotropic scattering model and $\psi$ is $83^{+4}_{-3}$ degrees, indicating that the diffraction pattern is a filament oriented parallel to the direction of declination of this pulsar.

Although the posterior distributions of $V_{\rm IISM, \delta}$ and $A_{r}$ for PSR~J0636+5128 do not converge well in the mildly
anisotropic scattering model, the scintillation velocities $V_{\rm{ISS, M}}$ and $V_{\rm{ISS, P}}$ are well matched with a reduced $\chi^{2}$ of 1.91 for the differences. Due to the negligible fluctuation of scintillation parameters caused by the pulsar orbital velocity and the absence of the epoch of periastron passage $T_{0}$, we neglected the pulsar orbital velocity when calculating $V_{\rm eff}$. As a result, we were unable to unambiguously constrain the scattering screen parameters and hence cannot entirely exclude the mildly anisotropic scattering model. This can be examined with well-resolved scintillation arcs, requiring longer observations and finer frequency resolution than we had in this work.

\subsection{Orbital angles $i$ and $\Omega_{\rm asc}$}
Fonseca et al. (2016) \cite{fpe+16} and Perera et al. (2019)  \cite{pdd+19} published the sine value of the inclination angle of PSR~J0613$-$0200 as 0.91$^{+0.05}_{-0.08}$ and 0.94$^{+0.02}_{-0.02}$, respectively.  
From our scintillation analysis, we can break that ambiguity by fitting both the sine and cosine of $i$. 
In our modelling, the inclination angle $i$ of PSR~J0613$-$0200 is 101$^{+8}_{-18}$ degrees, the $\sin(i)$ is 0.98$^{+0.02}_{-0.03}$ which is consistent with the timing result, but it points to $i \textgreater 90^{\circ}$. 
Additionally, we present the $\Omega_{\rm asc}$ of PSR~J0613$-$0200 as $-48^{+36}_{-9}$ degrees.
Unfortunately, the values of $i$ and $\Omega_{\rm asc}$ in our modelling are different with Main et al. (2023) \cite{mac+23}.

\section{Conclusions}
\label{sec:conclusion}
Assuming an anisotropic thin scattering screen model, we have modelled the annual variations of scintillation parameters using Equation~\ref{eqn:model}, allowing us to constrain the position, velocity and anisotropy of scattering screens of PSRs~J0613$-$0200 and J0636+5128. Comparing our results for PSR~J0613$-$0200 with Main et al. (2020) \cite{msa+20} and Main et al. (2023) \cite{mac+23}, our estimates for the distance of the scattering screen are consistent with the results in literature. 
Both scattering screens in this work are consistent with being associated with the Local Bubble shell. Since the Local Bubble shell appears as a closed surface \cite{zga+22}, the electron density fluctuations sharply increase at the shell along the propagation path of pulsar signals. Thus, for most nearby pulsars, the Local Bubble shell plays a substantial role in ISS \cite{bgr98,bg+02,bot+16}.

Early scintillation studies typically assumed isotropic scattering \cite{ric77}. Over the past two decades, there has been increasing evidence towards anisotropic and inhomogeneous \cite{wcv+23} scattering in many cases. 
Since the discovery of an extremely anisotropic distribution of images by Brisken et al. (2010) \cite{bmg+10}, the hypothesis of extremely anisotropic scattering has been frequently used in recent years. This hypothesis only requires velocities along the major axis of anisotropy, which results in fewer input parameters being necessary \cite{wdb+09}. Extremely anisotropic scintillation is supported in a few cases, e.g., sources with sharp inverted arclets (PSRs~B0834+06, B1508+55 and B0450$-$18 \cite{hsa+05,rsz+21,smw+22}). 
From our modelling, the scattering screens are both evidently anisotropic, one is even extremely anisotropic.

Additionally, we have presented $\Omega_{\rm asc}$ and $i$ for PSR~J0613$-$0200. The longitude of ascending node $\Omega_{\rm asc}$ is presented for the first time.  The calculation of the pulsar orbital velocity requires the introduction of both sine and cosine functions of $i$, which enable the determination of $i = 101^{+8}_{-18}~(\textgreater\,90^{\circ})$, thus resolving the ambiguity which timing was unable to resolve. 

\Acknowledgements{We thank Bill Coles for many useful discussions and suggestions. This work is supported by the National Natural Science Foundation of China (Grant No. 12003047), the Major Science and Technology Program of Xinjiang Uygur Autonomous Region (No. 2022A03013-2) and Natural Science Foundation of Xinjiiang Uygur Autonomous Region (No. 2022D01D85). 
JPWV acknowledges support by the Deutsche Forschungsgemeinschaft (DFG) through the Heisenberg programme (Project No. 433075039).
Part of this work is based on observations with the 100-m radio telescope of the Max-Planck-Institut f\"ur Radioastronomie (MPIfR) at Effelsberg in Germany. 
Pulsar research at the Jodrell Bank Centre for Astrophysics and the observations using the Lovell Telescope are supported by a consolidated grant from the STFC in the UK. 
The Nan\c{c}ay Radio Observatory is operated by the Paris Observatory, associated with the French Centre National de la Recherche Scientifique (CNRS). We acknowledge financial support from ``Programme National de Cosmologie et Galaxies'' (PNCG) of CNRS/INSU, France. }
%------------------------------------------------------------------------------------------------------------------------

%%%%%%%%%%%%%%%%%%%%%%%%%%%%%%%%%%%%%%%%%%%%%%%%%%%%%%%

%%%%%%%%%%%%%%%%%%%%%%%%%%%%%%%%%%%%%%%%%%%%%%%%%%%%%%%

\end{multicols}
\end{document}